# The Effect of Heat Loss During the Early Stages of Flame Propagation and Tulip Flame Formation

**Mikhail A. Liberman[a*] and Chengeng Qian [b]**

[a] *Nordita, KTH Royal Institute of Technology and Stockholm University, Hannes Alfvéns väg 12, 114 21 Stockholm, Sweden*

[b] *Aviation Key Laboratory of Science and Technology on High Speed and High Reynolds Number, Shenyang Key Laboratory of Computational Fluid Dynamics, Aerodynamics Research Institute, AVIC Shenyang 110034, China*

## ABSTRACT

The dynamics of premixed flames propagating in two-dimensional and cylindrical channels are investigated using direct numerical simulations of the multicomponent, fully compressible reactive Navier–Stokes equations coupled with conductive heat transfer within the channel walls. The simulations employ a high-order numerical method, detailed multistep chemical kinetics, and transport models for stoichiometric hydrogen–air combustion. Adaptive mesh refinement (AMR) is used to accurately resolve the flame structure, pressure waves, and their interactions. The influence of wall heat losses on the early stages of flame propagation is examined for channels with different aspect ratios, with particular emphasis on tulip flame formation and its subsequent transition to distorted tulip structures. Heat losses are modeled by accounting for convective heat transfer from the hot combustion products to the inner wall surface, thermal conduction through the wall, and convective and radiative heat losses from the outer wall surface to the surroundings. The results are compared with corresponding simulations under adiabatic wall boundary conditions. The simulations reproduce the principal features of tulip flame formation observed experimentally, highlighting the combined influence of wall heat losses and geometric confinement on flame dynamics in confined channels.

*Corresponding author: mliber@nordita.org (Michael Liberman)
Phone: +46855378444 (office)/ +46707692513 (mob)

## 1. Introduction

Flame propagation in channels is of great interest to study the fundamental features of the process in practical combustion devices, and it is a basic framework for developing and validating numerical and analytical models used to investigate a wide range of combustion-related engineering problems, including explosion safety, combustion in internal combustion engines, and pollutant emissions. The early stages of flame propagation are particularly important because they govern the initial acceleration or deceleration of the flame and thereby influence subsequent combustion regimes.

It has long been known that a flame ignited near the closed end of a tube and propagating toward the opposite closed or open end initially accelerates, exhibiting a convex flame front with the tip pointing forward. As the flame skirt approaches and eventually reaches the tube walls, the flame decelerates and the flame front rapidly changes from a convex to a concave shape, with the tip pointing backward. This phenomenon was first observed by Mallard and Le Chatelier [1] in 1883, photographed by Ellis [2] in 1928, and later termed the tulip flame (TF) [3]. Subsequent experimental studies demonstrated that tulip flame formation is remarkably robust across a wide range of combustible mixtures and downstream conditions, including open or closed tube ends, tube geometries, and wall roughness [4–12]. These characteristics establish the tulip flame as one of the fundamental phenomena in combustion science.

A convincing explanation of the tulip flame formation has remained elusive for a long time. Various physical processes, including Darrieus–Landau (DL) and Rayleigh–Taylor (RT) flame instabilities and flame interaction with vortices in the burnt gas, have been hyposized to explain TF formation [10–24]. Only recently, it was shown [25] that the tulip flame formation is a purely hydrodynamic phenomenon caused by a flame-generated rarefaction wave during the deceleration phase. It was demonstrated in [26–28] that flame-front inversion induced by rarefaction waves occurs much faster than the characteristic timescales of intrinsic flame instabilities suggesting that DL, thermal-diffusive, and RT instabilities are not the processes governing tulip flame formation, in agreement with experiments [29]. The subsequent evolution of the tulip flame and the formation of a distorted tulip flame (DTF) are strongly influenced by the intensity of pressure waves generated during the initial flame-acceleration stage.

In confined configurations, heat loss is unavoidable and can affect flame velocity, acceleration, structure, and subsequent combustion regimes such as the deflagration-to-detonation transition (DDT). Heat loss is therefore widely recognized as a key factor influencing flame propagation in tubes. Following the pioneering work of Zel'dovich [30], comprehensive studies of heat loss in narrow tubes [31–37] were conducted in relation to flammability limits and the quenching diameter, below which a flame cannot propagate because heat losses to the walls exceed the heat released by chemical reactions.

The development of hydrocarbon-fueled micro-burners [38, 39] further advanced studies of heat-loss in narrow tubes near the flammability limit. Accounting for realistic heat losses in narrow tubes and micro-burners is essential because their large surface-area-to-volume ratio enhances thermal interaction between hot combustion products and the tube wall. Detailed two- and three-dimensional simulations [40–45] have shown good agreement with experimental observations. These studies considered a premixed laminar flame stabilized in a narrow channel by a gas mixture with a flow rate that matched the flame's burning velocity. The effect of heat losses on the tulip flame formation and propagation was investigated in [46] assuming walls as thermally thin structures, which requires that the conductive heat transfer dominates the convective one, so that the heat is transferred inside the solid wall faster than it transferred by convection at to the wall boundary. Of interest is LES of explosions in venting chamber [47], where the boundary conditions on the walls are determined by the fact that walls do not have enough time to heat up, so that their temperature remains equal to the initial temperature.

The effect of heat loss during the initial stage of flame propagation in narrow tubes, $D < 10L_f$, was commonly studied using simplified approaches, such as single-step chemical kinetics and isothermal wall boundary conditions [48–56]. However, these models cannot accurately capture the full effects of heat loss on flame dynamics, resulting in discrepancies between numerical predictions and experiments. It is known [57, 58] that detailed chemical models reveal complex physical processes that are not captured by single-step chemistry. Likewise, isothermal wall boundary conditions do not adequately represent thermal interactions between high-temperature combustion products and tube walls.

The present work presents a comprehensive numerical investigation of heat-loss effects on tulip and distorted-tulip flame formation using simulations of the fully compressible Navier–Stokes equations with detailed multistep chemistry and convective heat transfer. Unlike previous studies, which employed isothermal wall boundary conditions, the present work accounts for convective heat transfer from combustion products to the wall inner surface, conductive heat propagation to the outer wall surface, and convective and radiative heat losses from the outer wall surface to the surroundings.

## 2. Numerical models and physical parameters

The system of interest, shown in Figure 1, is a two-dimensional or cylindrical channel of inner width (diameter) $D_{in} = 10mm$, with quartz or copper wall of width $d = 2mm$.

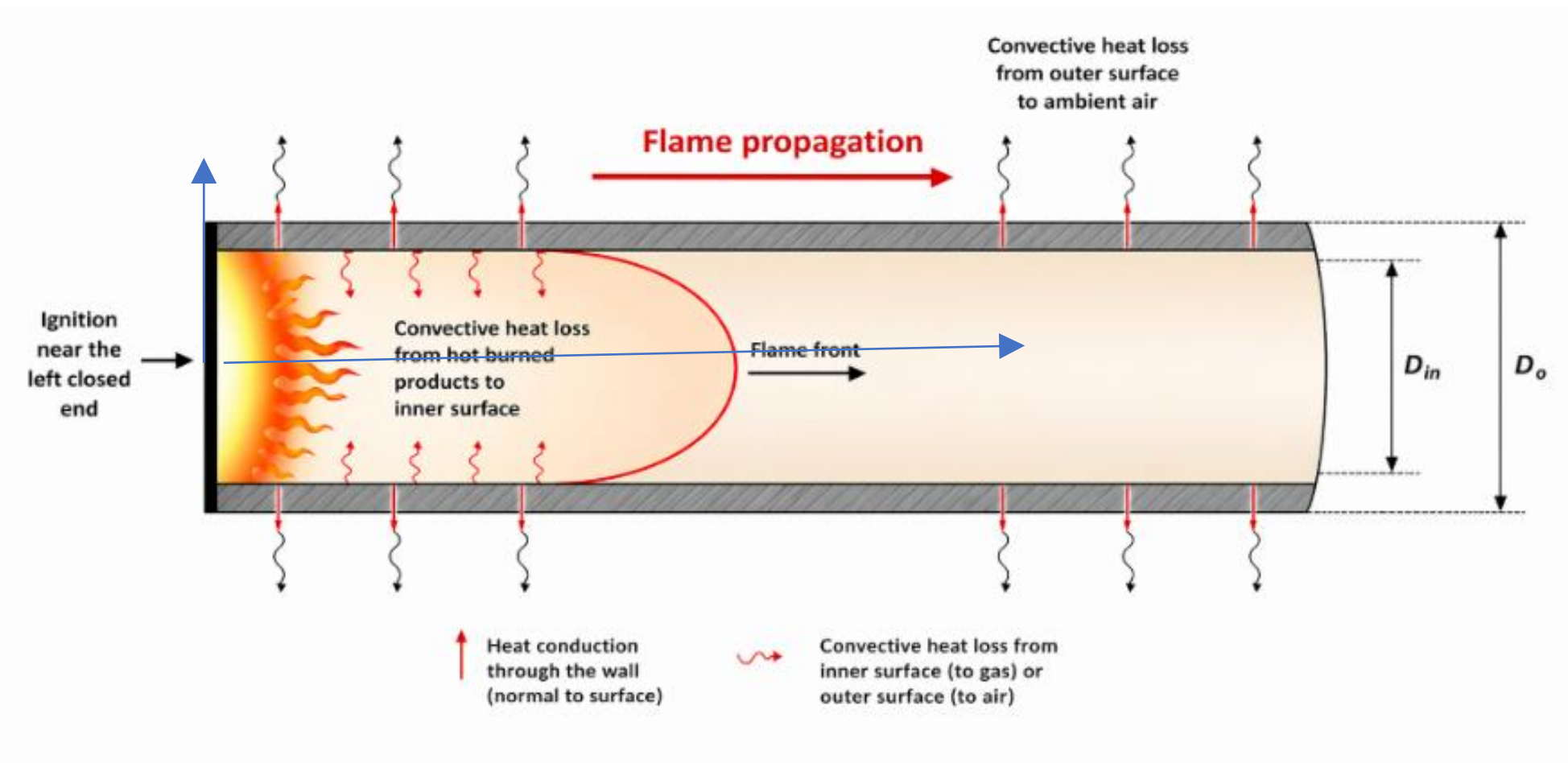

Fig. 1. Schematic of the 2D channel configuration: $D_0$ is outer width, $D_{in}$ is inner width, $D_0 = D_{in} + 2d = 14\,\text{mm}$, wall thickness $d = 2mm$.

A premixed flame is initiated at the closed left end of the channel by imposing a semicircular ignition kernel of hot burned gas with a radius of 1.5 mm. The kernel is centered on the channel centerline (tube axis) and attached to the left end wall such that its diameter lies in the plane of the wall. Following ignition, the flame propagates toward the opposite end of the channel, which may be either closed or open.

### *2.1. Governing equations*

The two-dimentional reactive Navier–Stokes equations are solved in their fully compressible form for gas phase of stoichiometric mixture for hydrogen-air including heat conductivity, multicomponent diffusion, viscosity and detailed multistep chemical kinetics of $H_2$/air. The continuity, momentum, energy, species, and equation of state equations for the reacting flow are:

$$\frac{\partial \rho}{\partial t} + \frac{\partial (\rho u_i)}{\partial x_i} = 0, \tag{1}$$

$$\frac{\partial (\rho u_i)}{\partial t} + \frac{\partial (P\delta_{ij} + \rho {u_i}^2)}{\partial x_j} = \frac{\partial \tau_{ij}}{\partial x_j}, \tag{2}$$

$$\frac{\partial (\rho E)}{\partial t} + \frac{\partial \left[ (\rho E + P) u_i \right]}{\partial x_i} = \frac{\partial (\tau_{ij} u_i)}{\partial x_i} - \frac{\partial q_i}{\partial x_i}, \tag{3}$$

$$\frac{\partial \rho Y_k}{\partial t} + \frac{\partial \rho u_i Y_k}{\partial x_i} = \frac{\partial}{\partial x_i}\left( \rho V_{ik} Y_k \right) + \dot{\omega}_k, \tag{4}$$

$$\sigma_{ij} = \frac{4}{3}\mu \left( \frac{\partial u_i}{\partial x_j} + \frac{\partial u_j}{\partial x_i} \right) - \frac{2}{3}\mu \frac{\partial u_i}{\partial x_i}\delta_{ij} \tag{5}$$

$$q_i = -\lambda_g \frac{\partial T}{\partial x_i} \tag{6}$$

The gas mixture is assumed to be an ideal gas with a constant ratio of specific heats $\gamma = c_P / c_V$

$$P = \rho R_B T \left( \sum_{i=1}^{N_s} \frac{Y_i}{W_i} \right). \tag{7}$$

The total energy is

$$E = \frac{P}{\rho(\gamma - 1)} + \frac{1}{2}(u_x^2 + u_y^2) \tag{8}$$

Here $\rho$, $u_i$, $T$, $P$, $E$, $\tau_{ij}$, $q_i$ are density, velocity components, temperature, pressure, total energy, the Newtonian viscosity stress tensor and heat flux. $R_B = 8.3144621 \text{J/mol K}$ is the universal gas constant, $Y_i, W_i, V_{i,j}$ are the mass fraction, molar mass and diffusion velocity of species $i$ ($i = 1,...,N_S$). The viscosity and thermal conductivity of species and mixture are calculated using the Chapman-Enskog expression [59] and a semi-empirical Wilke formula [60]. The x-axis is directed along the channel walls; $y = 0$ is on the channel centerline (Fig. 1).

The heat transfer in the solid walls is described by the thermal conduction equation with appropriate boundary conditions

$$\rho_W C_W \frac{\partial T_W}{\partial t} = \frac{\partial}{\partial x}\left( \kappa_W \frac{\partial T_W}{\partial x} \right) + \frac{\partial}{\partial y}\left( \kappa_W \frac{\partial T_W}{\partial y} \right), \tag{9}$$

where $\kappa_W$, $T_W$, $\rho_w$, $C_W$ are thermal conductivity, temperature, density and heat capacity of the solid wall.

### *2.2. Chemical model and initial parameters*

The reaction rate of species $i$ is determined as

$$\dot{\omega}_k = W_i \sum_{j=1}^{N_r} (v''_{jk} - v'_{jk}) \cdot \left( k_{f,j} \prod_{k=1}^{N_s} \left( \frac{\rho Y_k}{W_k} \right)^{v'_{jk}} - k_{b,j} \prod_{k=1}^{N_s} \left( \frac{\rho Y_k}{W_k} \right)^{v''_{jk}} \right) \tag{10}$$

where $v'_{jk}$ and $v''_{jk}$ are stoichiometric coefficients of species $k$ of the reactant and product sides of reaction $j$; $k_{f,j}$ and $k_{b,j}$ are the forward and backward rates of reaction j, which are calculated by Arrhenius equation.

A detailed chemical kinetic model for a stoichiometric hydrogen/air mixture consisting of 21 reactions and 9 species, developed by Kéromnès et al. [61] was implemented and used in the simulations. The parameters used in simulations are shown in Tables 1.

**Table 1.** Parameters for simulating stoichiometric hydrogen–air and walls

| | | |
|---|---|---|
| Initial pressure | $P_0$ | 1.0 atm |
| Initial temperature | $T_0$ | 298 K |
| Initial density | $\rho_0$ | $8.5 \cdot 10^{-4}\ \mathrm{g/cm^3}$ |
| Laminar flame velocity | $U_f$ | 2.43 m/s |
| Laminar flame thickness | $L_f$ | 0.0325 cm |
| Adiabatic flame temperature | $T_b$ | 2350 K |
| Expansion coefficient $\rho_u / \rho_b$ | $\Theta$ | 8.34 |
| Specific heat ratio | $\gamma = C_P / C_V$ | 1.399 |
| Sound speed | $a_s$ | $408.77\mathrm{m/s}$ |
| Thermal conductivity of quartz | $\kappa_{wQ}$ | 1.4 W/(m·K) |
| Density of quartz | $\rho_Q$ | 2650kg/m$^3$ |
| Specific heat capacity of quartz | $C_{p,Q}$ | 743 J/(kg·K) |
| Thermal conductivity of copper | $\kappa_{wCo}$ | 400 W/(m·K) |
| Density of copper | $\rho_{C0}$ | 8960kg/m$^3$ |
| Specific heat capacity of copper | $C_{PCo}$ | 385 J/(kg·K) |

As noted above, the results obtained using a detailed chemical mechanism differ significantly from those obtained with a single-step chemical model. Figures A1(a,b) in the Appendix illustrate the differences between simulations performed using a single-step chemical model and those employing detailed chemistry under adiabatic boundary conditions. Similarly, Figures A2(a,b) compare the results of simulations using a single-step chemical model with those obtained using detailed chemistry while accounting for heat losses.

### *2.3. Numerical method*

The fifth-order weighted essentially non-oscillatory (WENO) finite-difference method [62] is employed to solve the two-dimensional, time-dependent, reactive compressible Navier–Stokes equations (1–8) for the gas-phase within the channel, accounting for molecular diffusion, thermal conduction, and viscosity. A key advantage of the WENO finite-difference scheme is its ability to achieve arbitrarily high-order accuracy in smooth regions while robustly capturing sharp discontinuities. To ensure numerical conservation, the nonlinear diffusion terms are discretized using a fourth-order conservative central-difference scheme [63]. Time integration is performed using a third-order strong-stability-preserving Runge–Kutta (SSP-RK) method [64].

Accurate modeling of reactive flows requires sufficient resolution of the internal flame structure. Because the laminar flame thickness decreases as pressure increases, a finer grid resolution is required at the maximum pressure. Grid-independence studies reported in our previous work [25–28] determined the necessary resolution by systematically varying the grid spacing, $\Delta x$, to eliminate numerical artifacts. Based on these findings, a uniform mesh with grid spacing $\Delta x$ was employed, providing approximately 28 grid points across the flame thickness under the initial conditions and 14 grid points at the maximum pressure in a tube closed at both ends.

### *2.4 Boundary conditions*

Symmetry conditions are imposed at the channel centerline (midplane) between the sidewalls. This assumption is justified by the robust nature of tulip flame formation, which has been shown to persist despite variations in mixture composition and downstream boundary conditions, as demonstrated in numerous experimental studies.

$$\frac{\partial u_x}{\partial y} = \frac{\partial u_y}{\partial y} = \frac{\partial T}{\partial y} = \frac{\partial Y_k}{\partial y} = 0 \text{ at } (x, y = 0). \quad (11)$$

Depending on the experimental conditions, gravitational buoyancy may introduce asymmetry into the tulip flame shape, thereby affecting its overall structure and appearance. However, the characteristic buoyancy timescale is generally much longer than the tulip flame formation timescale, even for low-reactivity methane–air mixtures. Therefore, gravitational effects are neglected in the present study. The walls are assumed to be chemically inert, and heterogeneous reactions on the inner wall surface are not considered (see, e.g., [65, 66]).

No-slip and impermeable boundary conditions are imposed at the wall

$$\vec{u}\big|_{y=\pm D_i/2} = 0\,,\ Y_k\big|_{y=\pm D_i/2} = 0\,. \qquad (12)$$

Under no-slip conditions a velocity boundary layer adjacent to the wall develops. Likewise, when the burned gas comes into contact with the inner surface of the channel wall, a thermal boundary layer is established [67]. Consequently, the heat boundary condition on the inner wall surface is convective heat transfer, which is described by Newton's cooling law [68, 69].

$$\left(\lambda_g \nabla T_g \cdot \vec{n}_i + (\hat{\tau}\cdot\vec{u})_g \cdot \vec{n}_i\right)\Big|_{y=D_{in}/2} = h_i (T_{g,0} - T_w)\,, \qquad (13)$$

where $h_i$ is the convective heat transfer coefficient in [W/m$^2$K], $T_{g,0}$ is the gas temperature outside the boundary layer (at the tube axis), $T_w$ is the temperature on the inner surface of the wall, $\lambda_g = 0.186 W/mK$ is the burnt gas thermal conduction. Because the area of the left closed end is much smaller than that of the channel sidewalls, heat losses to the closed end are negligible, and the adiabatic no-slip boundary condition is applied to the left closed end. Here the local heat flux density transmitted by the gas-phase through the inner wall surface is the sum of the diffusive heat flux density and the viscous stress power at the wall

$$q_i = (\lambda_g \nabla T_g + (\hat{\tau}\cdot\vec{u})_g)\cdot\vec{n}_e\,, \qquad (14)$$

where $\vec{n}_e$ is the normal unit vector to the channel's wall surface, $\hat{\tau} = \mu(\nabla\vec{u} + \nabla\vec{u}_\perp) - \frac{2}{3}\mu(\nabla\cdot\vec{u})\hat{I}$ is the viscous stress tensor, $\mu$ is the viscosity coefficient of the gas mixture and $\hat{I}$ is unite tensor. The effect of viscous stress heating with an accuracy of a numerical factor of order unity is $\Delta T / T_0 \approx M_b^2$, where $M_b = u_b / a_{sb}$ is the Mach number in the burnt gas flow. In the early stages, $M_b \le 10^{-2} \div 10^{-3}$, viscous heating is negligible, and Eq. (13) reduced to

$$\left(\lambda_g \nabla T_g \cdot \vec{n}_i\right)\Big|_{y=D_{in}/2} = h_i (T_{g,0} - T_w)\,. \qquad (13a)$$

Since the rate of heat propagation is determined by thermal diffusivity, $\chi = \kappa / \rho C$, heat propagates much slower in solids than in gases, $U_w / U_f \propto \rho_g C_g / \rho_w C_w \sim 10^{-5}$. Therefore, in equation (9), we can consider heat transfer through the wall only in the direction perpendicular (along the y-axis) to the wall surface, and Eq. (9) reduces to

$$\rho_W C_W \frac{\partial T_W}{\partial t} = \frac{\partial}{\partial y}\left( \kappa_W \frac{\partial T_W}{\partial y} \right). \qquad (9a)$$

***2.5 Convective heat transfer and Nusselt number***

The dimensionless quantity that characterizes convective heat transfer is the Nusselt number.

$$Nu = h_i L / \lambda_g , \qquad (15)$$

where $h_i$ is the heat transfer coefficient, $L$ is the characteristic length of the problem. From similarity arguments it follows that the Nusselt number is function of the Reynolds and Prandtl numbers only [67-69]. The average value of $h_i$ can be approximated as

$$\bar{h}_i = Nu \lambda_g / L . \qquad (16)$$

For the laminar flow inside the channel the average Nusselt number for horizontal plates can be estimated as [67 - 69]

$$\overline{Nu} = 0.664 \, \mathrm{Re}^{1/2} \, \mathrm{Pr}^{1/3} \qquad (17)$$

According to [70], experimental data for laminar flow in tubes can be better approximated by the equation

$$Nu = 1.86 \left( Pe \frac{D_{in}}{L} \right)^{1/3} \left( \frac{\mu_b}{\mu_w} \right)^{0.14} , \qquad (18)$$

where $Pe = Re \cdot Pr$ is Péclet number, $L / D_{in}$ is the aspect ratio. Both equations (17) and (18) provide fairly close values for the Nusselt number. For the burnt gas in a tube with aspect ratio $L / D_{in} = 6$, $Pe = Re \cdot Pr = 166 \cdot 0.7 = 116$, $\mu_b / \mu_w = 4.9$, we obtain

$$Nu = 1.86\left(Pe\frac{D_{in}}{L}\right)^{1/3}\left(\frac{\mu_b}{\mu_w}\right)^{0.14} = 6.25. \quad (19)$$

For this value of the Nusselt number the convective heat transfer coefficient is

$$h_i = Nu\,\lambda_g / D_{in} \approx 116\,\mathrm{W/m^2K}. \quad (20)$$

Taking the coefficient of thermal conduction for hydrogen/air combustion products, $\lambda_g(T_b) = 0.186\,\mathrm{W/mK}$, kinematic viscosity $\nu_b \simeq 6cm^2/s$, and maximum velocity in the burnt gas, $U_b \sim 10^3 cm/s$, we obtain the average value of the Nussel number

$$\overline{Nu} = 0.664\,\mathrm{Re}^{1/2}\,\mathrm{Pr}^{1/3} \approx 6.11. \quad (21)$$

Then, the average value of the convective heat transfer coefficient is very close to that given by Eq. (20):

$$\overline{h}_i = \overline{Nu}\,\lambda_g / D_{in} = 113.6\,\mathrm{W/m^2K}. \quad (22)$$

Constant heat transfer coefficients are considered as realistic for laminar flows in plane channels [68], with the Nusselt number based on hydraulic diameter, $Nu = h_i D_H / \lambda_g \approx 7.5$,

$$h_i \approx 7.5\lambda_g / D_{in} = 139.5\,\mathrm{W/m^2K}. \quad (22a)$$

Because different approaches give the values of the convective heat transfer coefficient that differ by $\pm 10\%$, we will use in simulations $h_i = 100\,\mathrm{W/m^2K}$. A high-resolution boundary layer is necessary in order to ensure the accuracy of the convective heat transfer model.

The boundary conditions at the external wall surface in contact with ambient air at normal temperature $T_0 = 300K$ are given by the convective and the radiative heat loss [69-70]

$$\mathbf{n}_{ext}\cdot(-\kappa_W \nabla T_W) = h_{ext}(T_{W,ext} - T_0) + \varepsilon\sigma(T_{W_{ext}}^4 - T_0^4), \quad (23)$$

where $\sigma = 5.670\times 10^{-8}\,\mathrm{Wm^{-2}K^{-4}}$ is the Stefan–Boltzmann constant. The external convective heat transfer coefficient is $h_{ext} = 5\,\mathrm{W/m^2}K$ for weak natural convection; the upper limit of natural convection is $h_{ext} = 20\,\mathrm{W/m^2}K$. The emissivity is $\varepsilon = 0.1 \div 0.3$ for the polished quartz wall surface $\varepsilon = 0.2$.

## 3. Two dimensional channel

### *3.1. 2D Channel with Both Ends Closed; Aspect Ratio* $\alpha = L / D_{in} = 6$

In this section we compare the results of modelling dynamics of hydrogen/air flame for adiabatic walls with the flame dynamics obtained accounting for heat loss to the walls for the flame propagating in a 2D channel with both ends closed and aspect ratio $\alpha = 6$. Fig. 2 shows the time evolution of the flame front velocities along the channel axis, $U_f(y = 0)$, and the flame front velocity close to sidewall, $U_f(y = 0.42cm)$. The solid lines in Fig. 2 correspond to calculations for adiabatic boundary conditions. The lines with squares and triangles represent calculations that account for the heat loss.

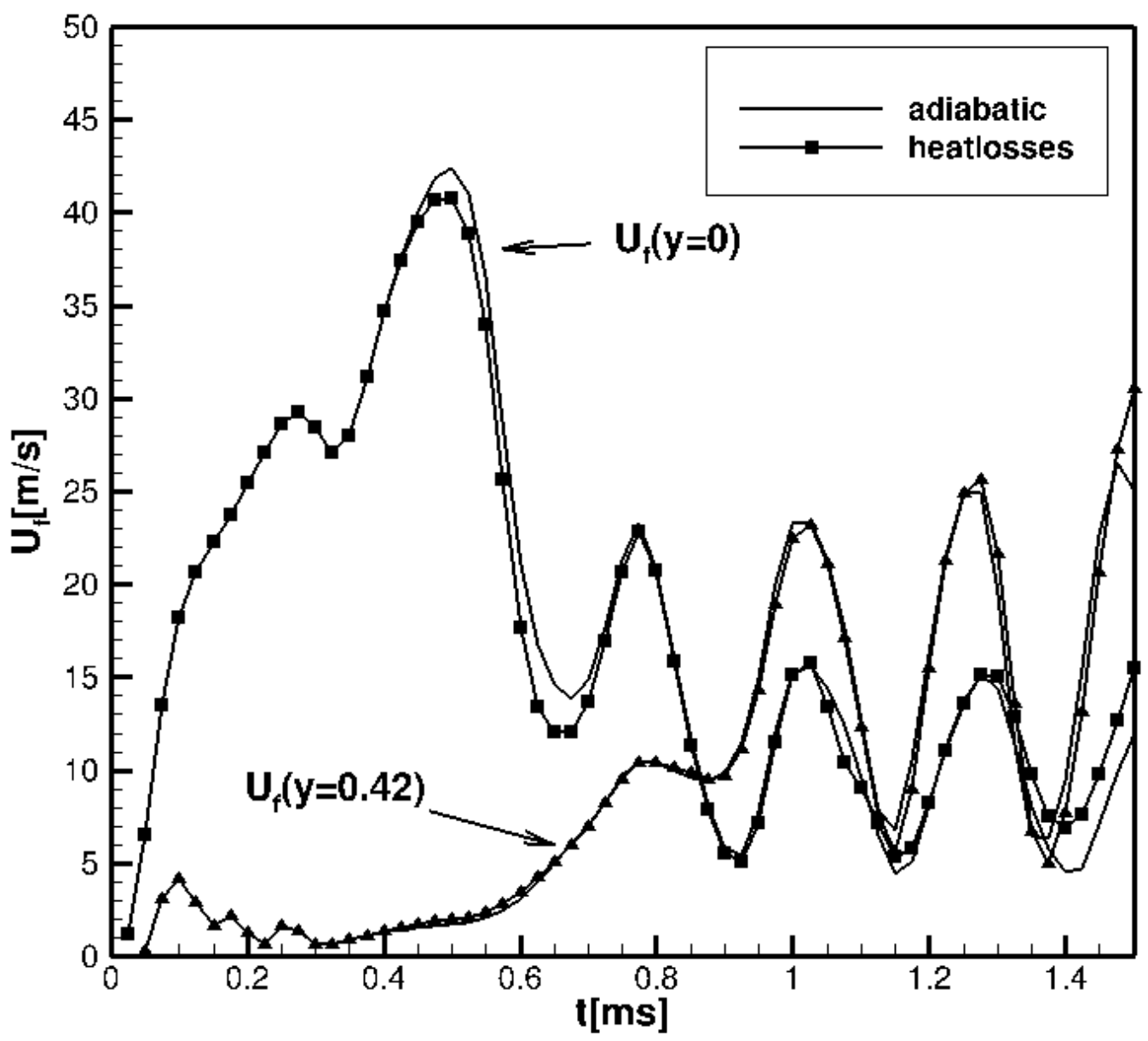

Fig. 2. Evolution of flame front velocity for adiabatic and heat loss boundary condition, $\alpha = 6$, the convective heat transfer coefficient: $h_i = 100\,\mathrm{W/m^2K}$, the thickness of quartz wall $d = 2mm$.

Figure 2 shows that the time at which the flame front begins to invert—that is, when the velocity of the flame front near the wall exceeds the velocity of the flame front along the channel axis—is the same for both adiabatic and heat loss boundary conditions. This is because only a small amount of heat is transferred from the combustion products during a short time, about 1ms. It is seen that due to heat loss the flame velocity along the axis, $U_f(y=0)$ only slightly decreases compared to $U_f(y=0)$ obtained in simulations with adiabatic boundary conditions.

Figures 3a and 3b show computed schlieren images and streamlines for $H_2$/air flame in the channel with both ends closed and aspect ratio $\alpha = 6$ obtained in simulations with adiabatic boundary conditions (Fig. 3a), and with heat loss boundary conditions (Fig. 3b).

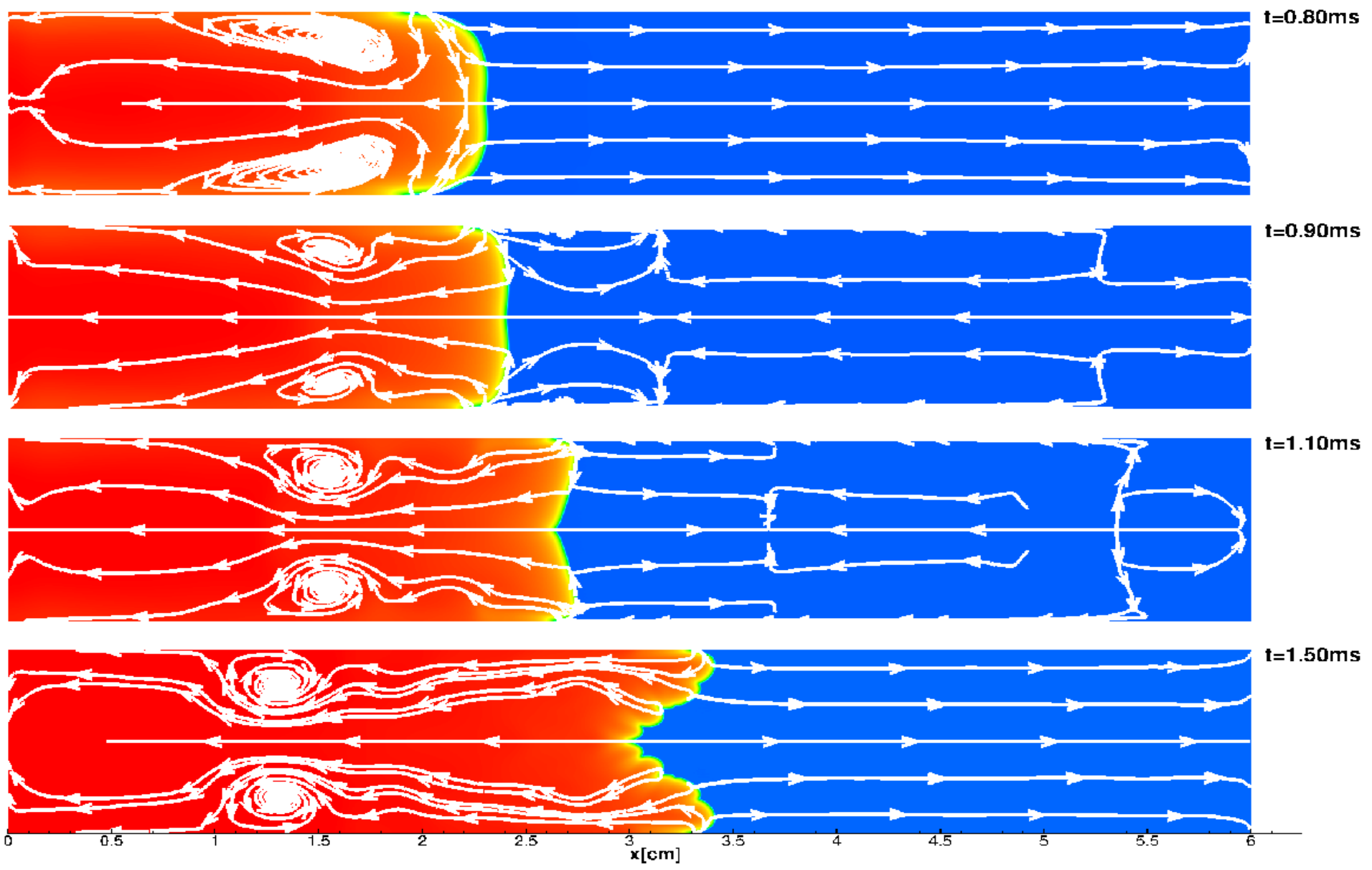

Fig. 3a. Computed schlieren images and streamlines for $H_2$/air flame in the channel $\alpha = 6$; the thickness of quartz wall $d = 2mm$, adiabatic boundary conditions.

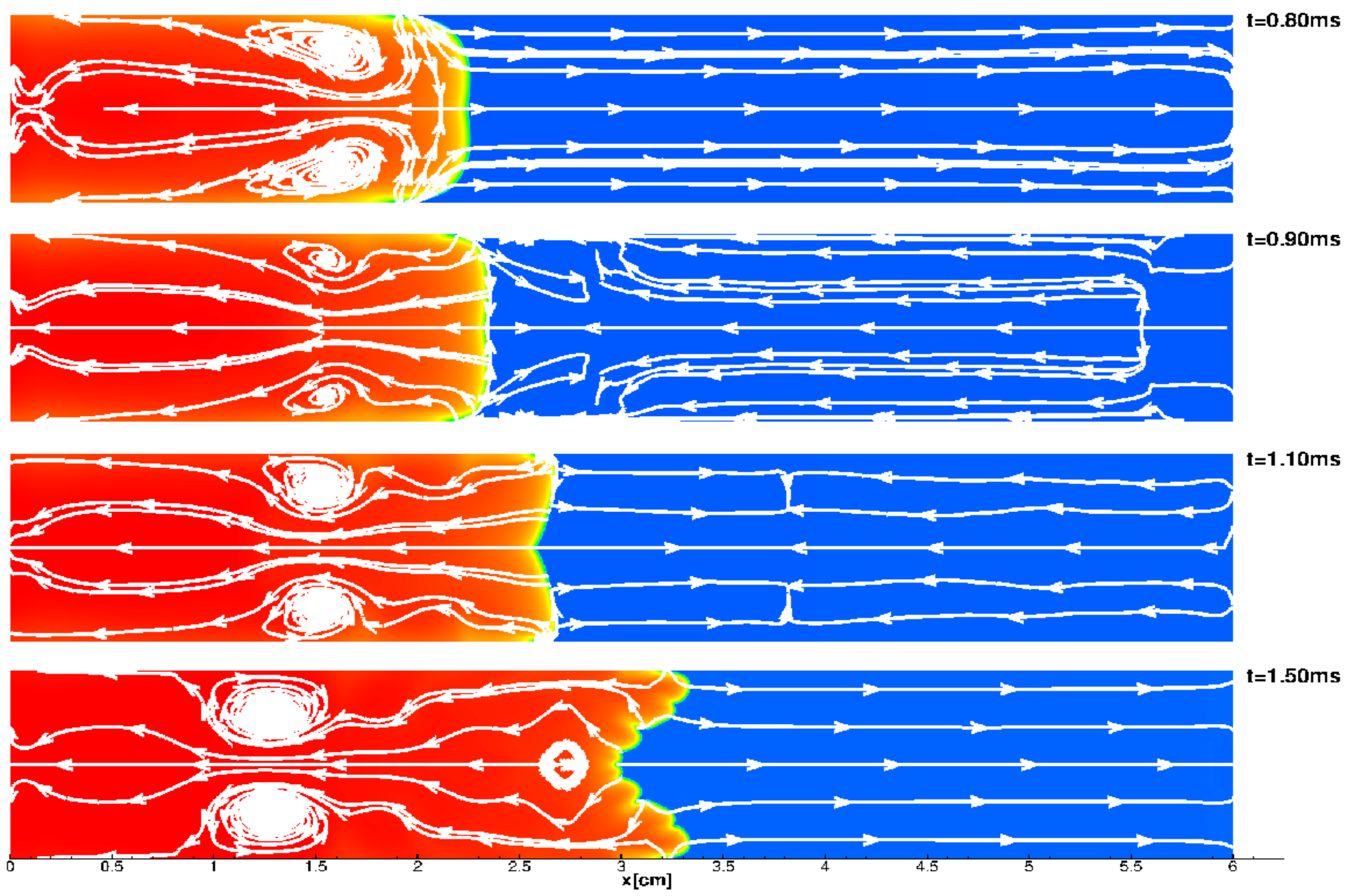

Fig. 3b. Computed schlieren images, temperature and streamlines, $H_2$/air flame in the channel $\alpha = 6$; heat loss: $h_i = 100\,\mathrm{W/m^2K}$, the thickness of quartz wall $d = 2\,mm$.

As noted above, the effect of heat loss remains small because of the relatively short propagation time. At the end of the finger-flame stage, when the flame-tip velocity $U_f(y=0)$ reaches its maximum, heat loss manifests itself as a small reduction in the flame-tip velocity, immediately before the onset of flame-front inversion at (t = 0.85) ms. At later times. t > 1.4 ms, after the tulip-flame petals have formed and become distorted by Rayleigh–Taylor instabilities [25–28], the flame-tip velocity in the heat-loss case slightly exceeds that obtained under adiabatic boundary conditions, as shown in Fig. 2.

### *3.2. Two dimensional channel with both ends closed, aspect ratio $L/D_{in}$=12*

As shown in Refs. [26, 27], the flame collisions with pressure waves reflected from the opposite closed end of the channel may enhance the effect of the rarefaction wave, thereby accelerating flame front inversion. In the longer channel, the flame undergoes fewer collisions with the reflected pressure waves, whose intensity is also weaker than in the shorter channel. As a result, flame front inversion occurs later in the longer channel. Moreover, the flame deceleration caused by these weaker pressure wave–flame interactions may be insufficient to trigger Rayleigh–Taylor instabilities and the subsequent formation of a distorted tulip flame. This behavior is evident in the calculated Schlieren images shown in Figures 4a and 4b.

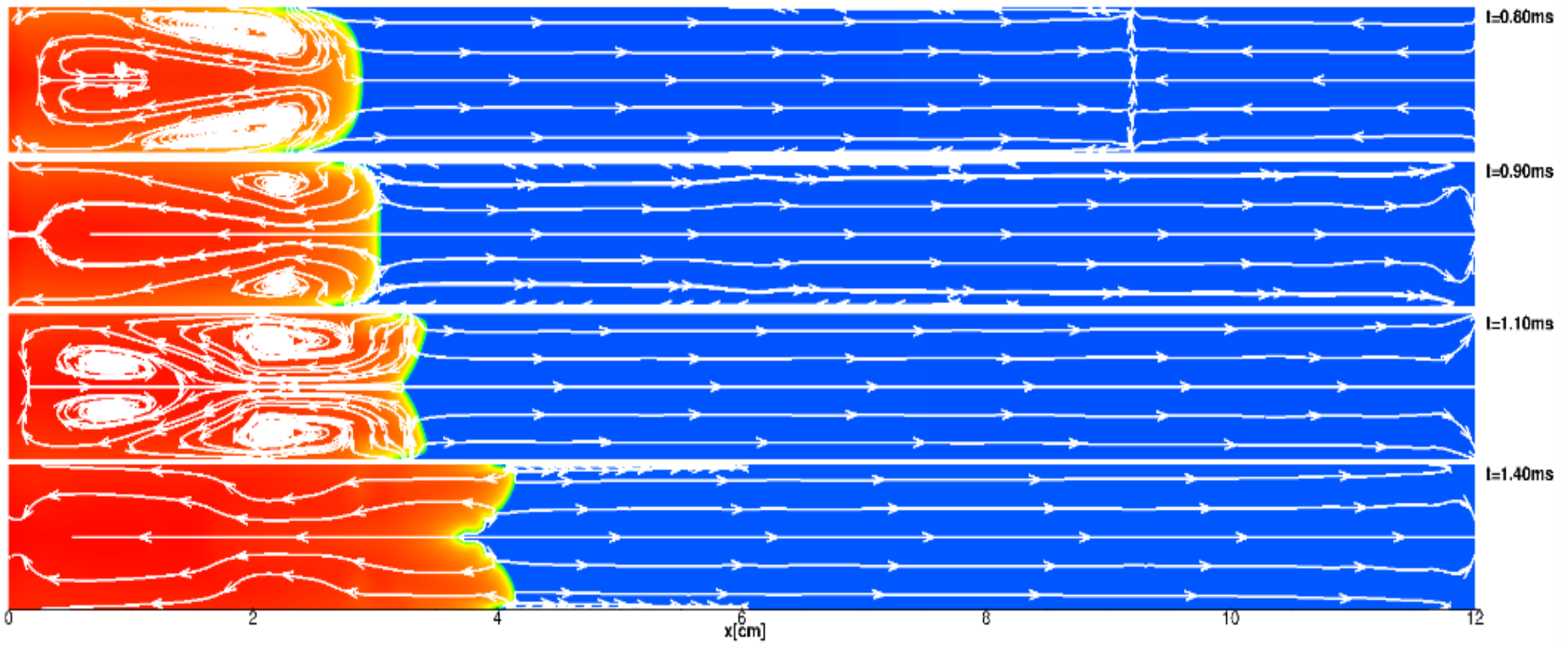

Fig. 4a. Computed schlieren images and streamlines: $H_2$/air flame; the channel length $L = 12cm$, $h_i = 100\,\mathrm{W/m^2K}$; quartz wall of thickness $d = 2\,mm$, adiabatic boundary conditions.

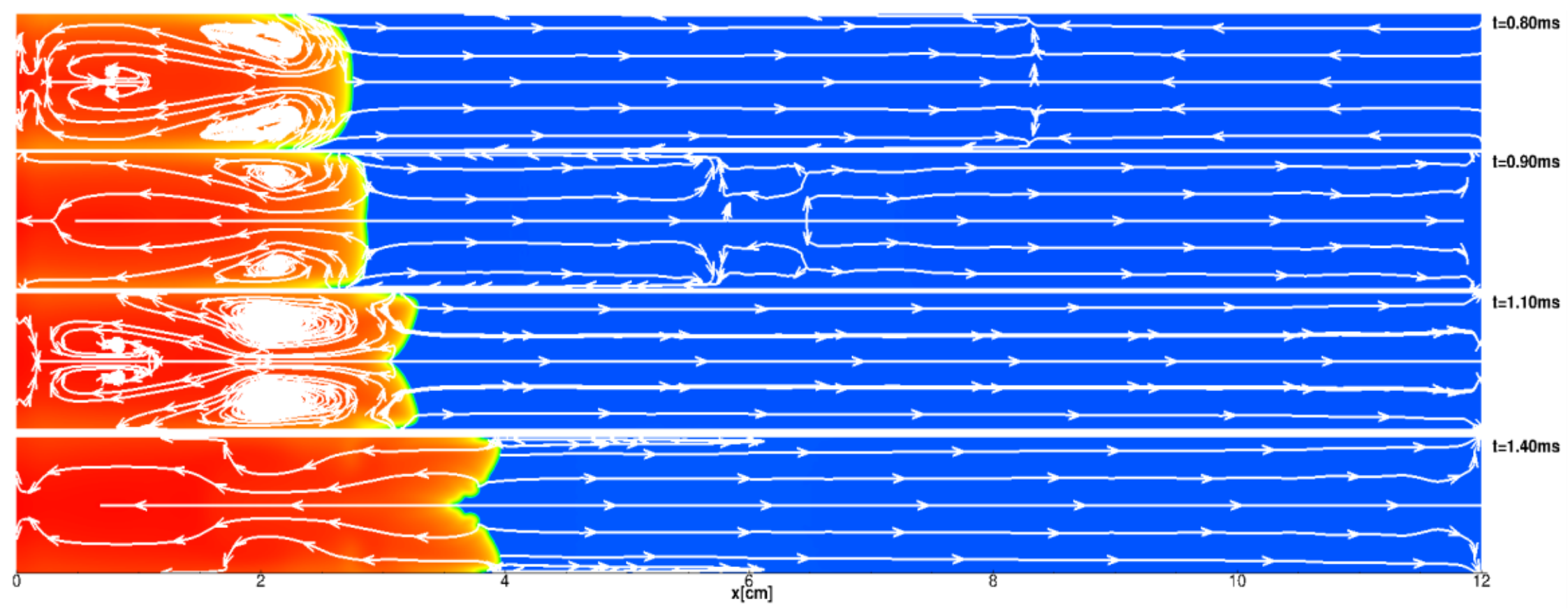

Fig. 4b. Computed schlieren images and streamlines for $H_2$/air flame. The channel length $L = 12cm$; heat losses: $h_i = 100\,\mathrm{W/m^2K}$, quartz wall of thickness $d = 2\,mm$.

The effect of heat loss is not readily discernible from the computed Schlieren images, as the differences between the adiabatic case (Fig. 4a) and the case with heat losses (Fig. 4b) are subtle. In contrast, Fig. 5 clearly reveals the influence of heat loss through the flame front velocity, which differs significantly between the adiabatic case (solid line) and the cases with heat losses (filled squares and triangles).

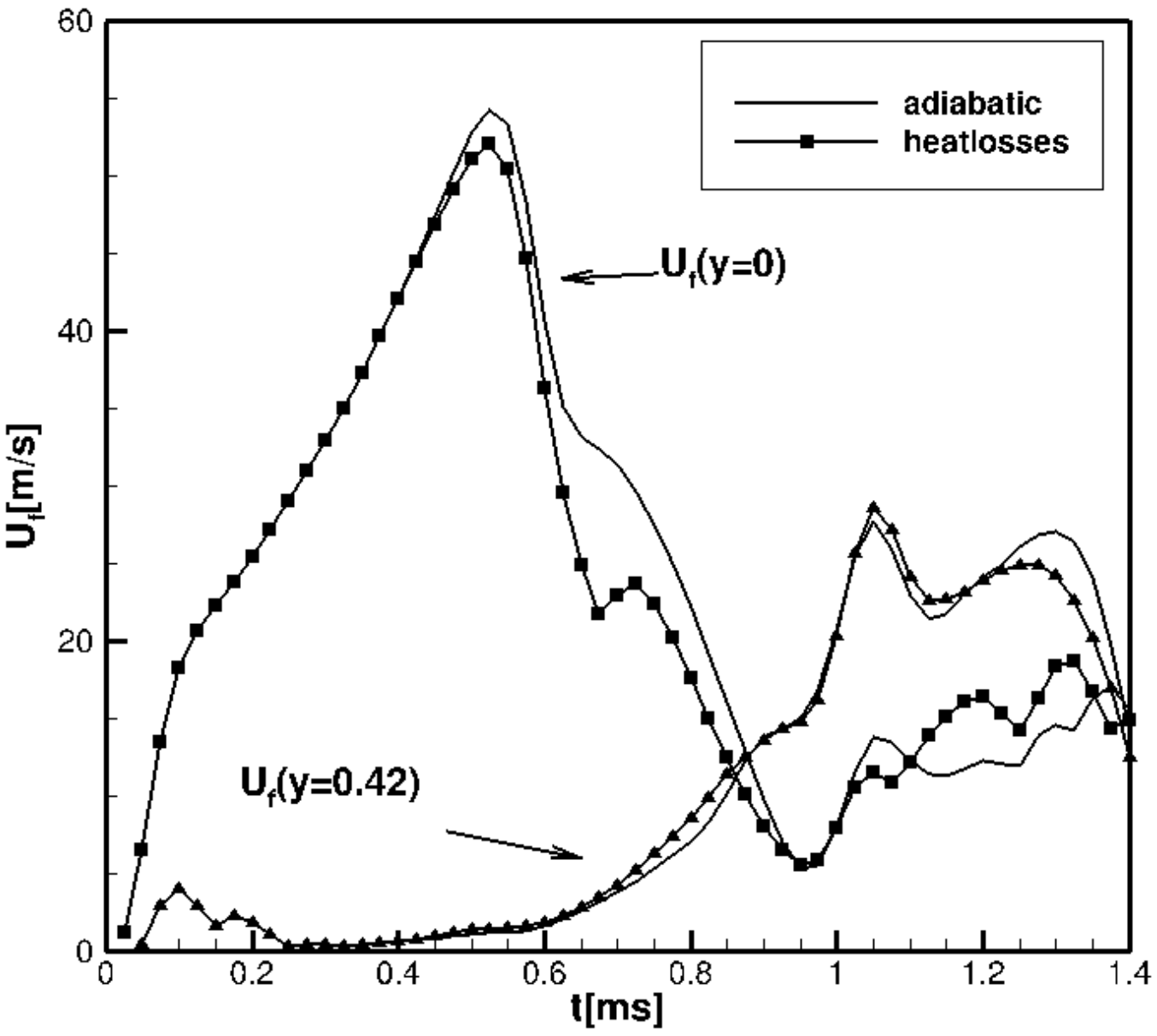

Fig. 5. Evolution of flame front velocities at the channel axis and near the wall calculated for adiabatic (solid lines) and heat loss (marked lines) boundary conditions. $H_2$/air flame, aspect ratio $L/D_{in} = 12$, heat losses: $h_i = 100\,\mathrm{W/m^2K}$, quartz wall thickness $d = 2\,mm$.

### *3.3 Effect of the wall thermal conductivity*

As discussed above, heat propagates much more slowly in solids than in gases because the thermal diffusivity of the wall material, $\chi_w = \kappa_w / \rho_w c_w$, is significantly lower than that of the gas mixture. Consequently, heat conduction within the wall can be assumed to occur only in the direction normal to the wall surface. Furthermore, because of the low thermal diffusivity of the wall material and the short duration of the early stage of flame propagation, during which the tulip and distorted tulip flames develop, the thermal penetration depth remains smaller than the wall thickness. As a result, the temperature of the outer wall surface remains essentially unchanged for sufficiently thick walls. Figure 6 shows the evolution of the flame front velocity along the tube axis, $y = 0$, and near the wall, at $y = 0.42cm$, for both adiabatic and heat-loss boundary conditions. The simulations presented in Fig. 6 were performed for a two-dimensional channel with both ends closed, an aspect ratio of $\alpha = 6$, and copper walls of thickness $d = 2\,mm$. It is worth noting that, even for copper walls, whose thermal conductivity is approximately 120 times higher than that of quartz, the heat does not reach the outer wall surface during the period considered.

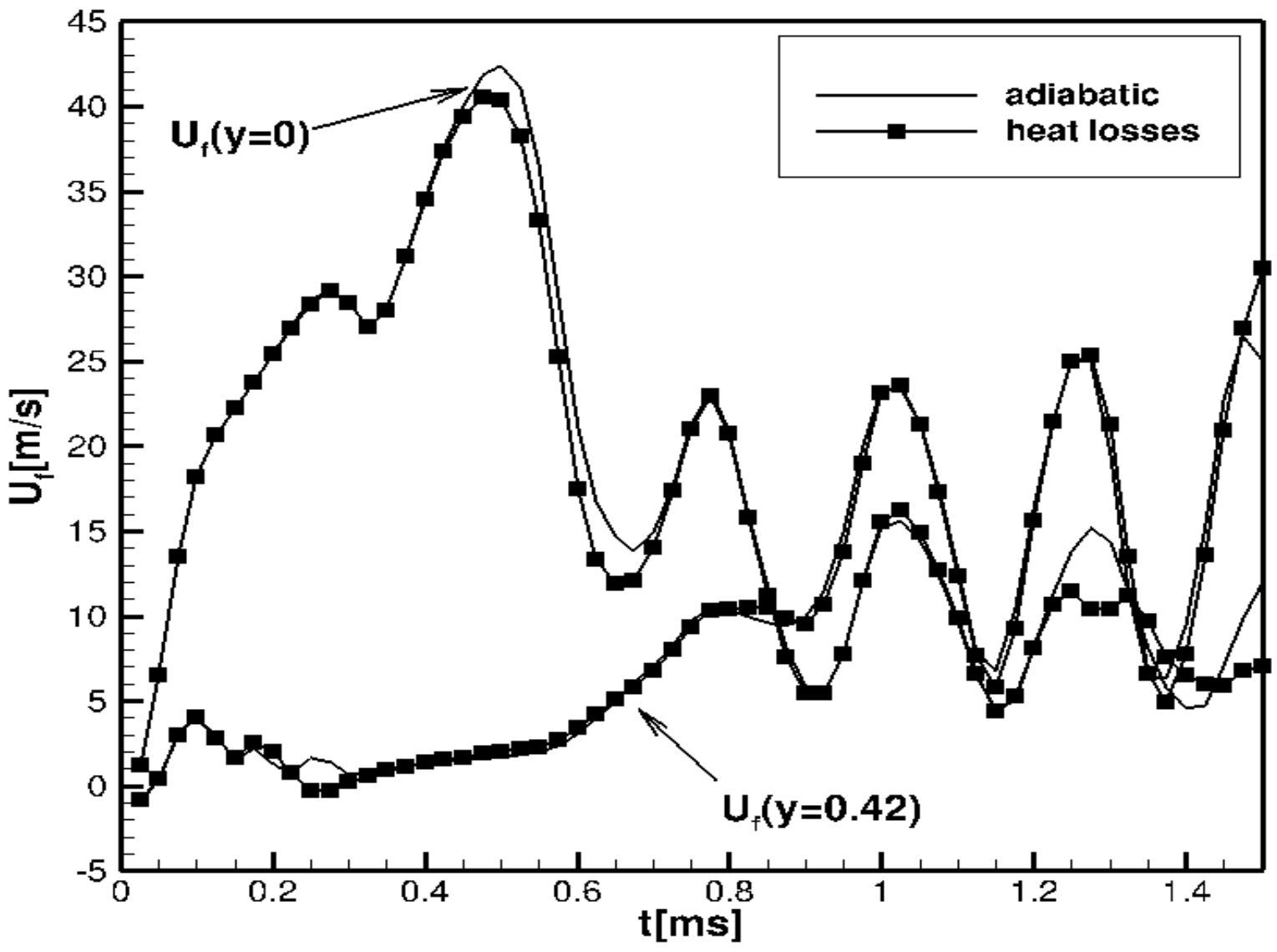

Fig. 6. Evolution of flame front velocity at the channel axis $y = 0$ and near the wall $y = 0.42cm$ calculated for adiabatic and heat-loss boundary condition; $\alpha = 6$, convective heat-loss: $h_i = 100\,\mathrm{W/m^2K}$, copper wall of thickness $d = 2\,mm$.

Consequently, the temperature of the outer wall surface remains nearly the same as the surroundings one for both quartz and coppe 2 mm thick wall. For quartz walls thinner than 0.1 mm and copper walls thinner than 1 mm, heat losses from the outer surface to the surroundings by convection and radiation become significant and must be incorporated through the boundary condition given by Eq. (23). Temperature profiles inside the quartz and copper walls at $t = 1.5\,\mathrm{ms}$, and the location $x = 1.5\,\mathrm{cm}$ are shown in Fig. 7.

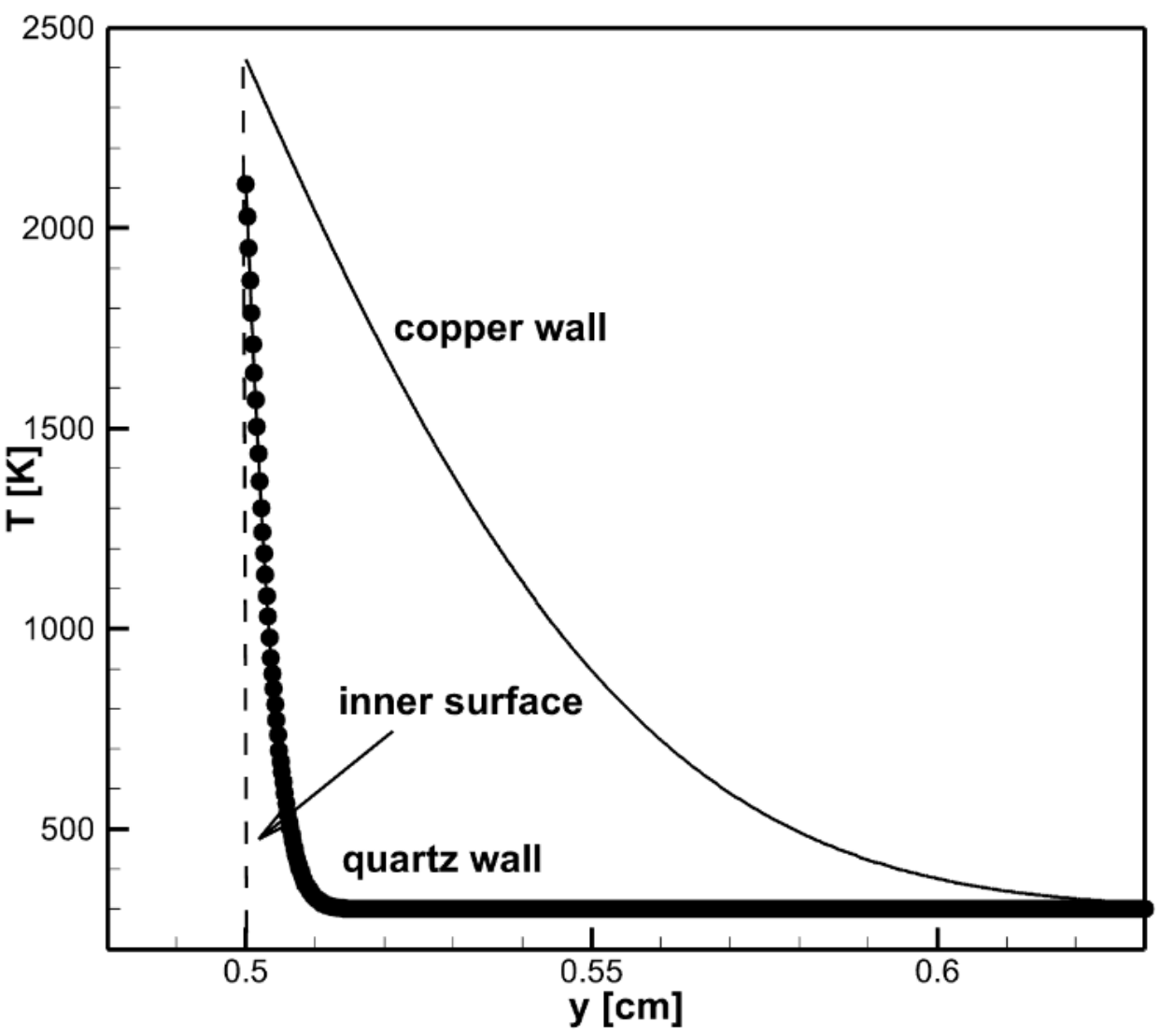

Fig. 7. Temperature profiles inside the channel wall at t=1.5ms calculated for quartz and copper walls. The inner surface of the wall is at 0.5cm, $d = 2mm$, $h_i = 100\,\mathrm{W/m^2K}$.

Figures 8(a,b) show the temporal evolution of the temperature profiles within the 2 mm thick quartz walls of a channel with both ends closed and an aspect ratio of $\alpha = 6$. The temperature distributions are presented at two locations: (a) near the left closed end, at $x = 0.1\,\mathrm{cm}$; and (b) farther from the closed end, at $x = 1.5\,\mathrm{cm}$. In both locations, the temperature of the inner wall surface increases with time as heat is transferred from the hot combustion products to the wall. The higher wall temperature observed near the closed end at t=0.8 ms probably results from the adiabatic boundary condition imposed at the end wall, which prevents heat loss through the boundary and leads to greater heat accumulation in this region.

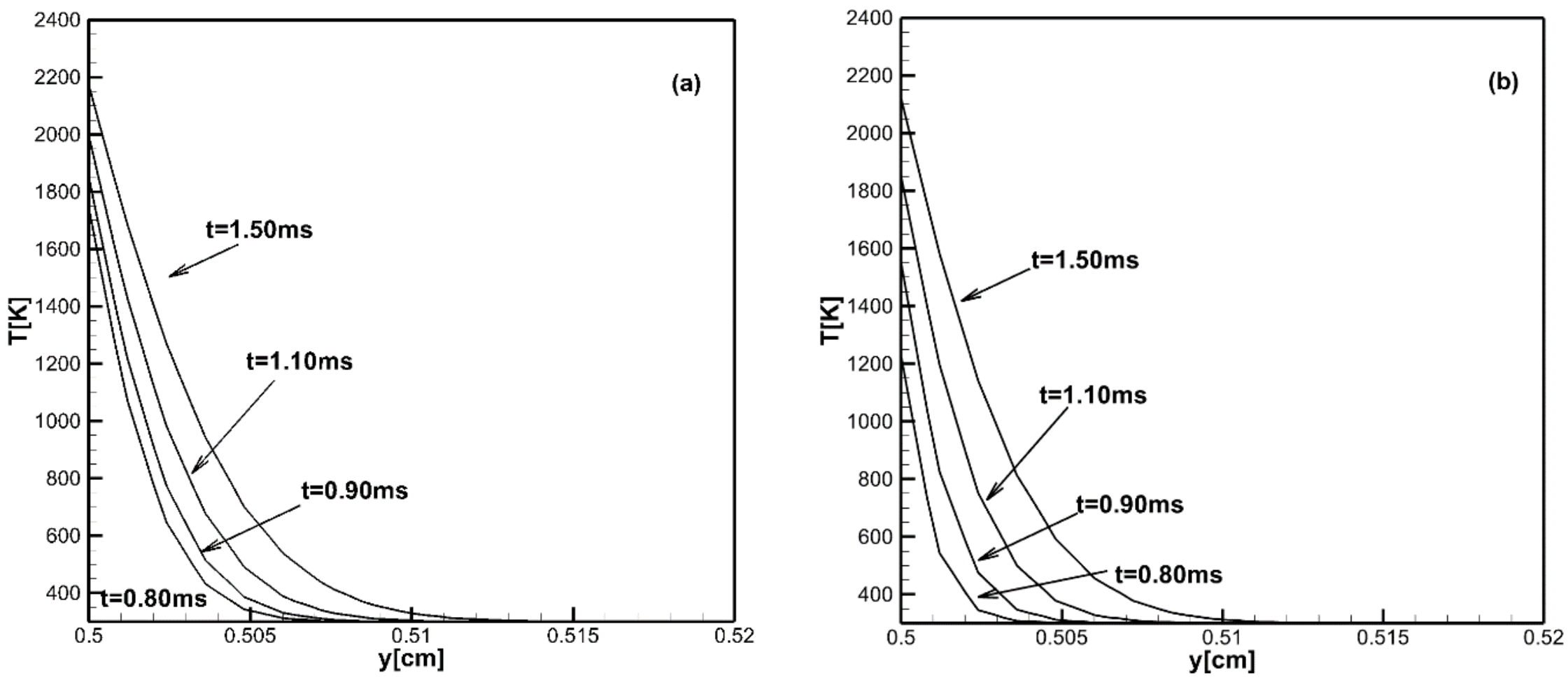

Fig. 8 (a,b). Calculated temperature profiles inside the quatz channel wall at different times; (a): near the left closed end, at $x = 0.1cm$, (b) at $x = 1.5cm$. $L/D = 6$, $d = 2mm$, $h_i = 100\,\mathrm{W/m^2K}$.

Figures 9(a, b) show similar the time evolution of the temperature profiles within the copper cwalls. Figure 9(a) corresponds to the location near the left closed end, at $x = 0.1\mathrm{cm}$, while Figure 9(b) corresponds to the location at $x = 1.5\,\mathrm{cm}$.

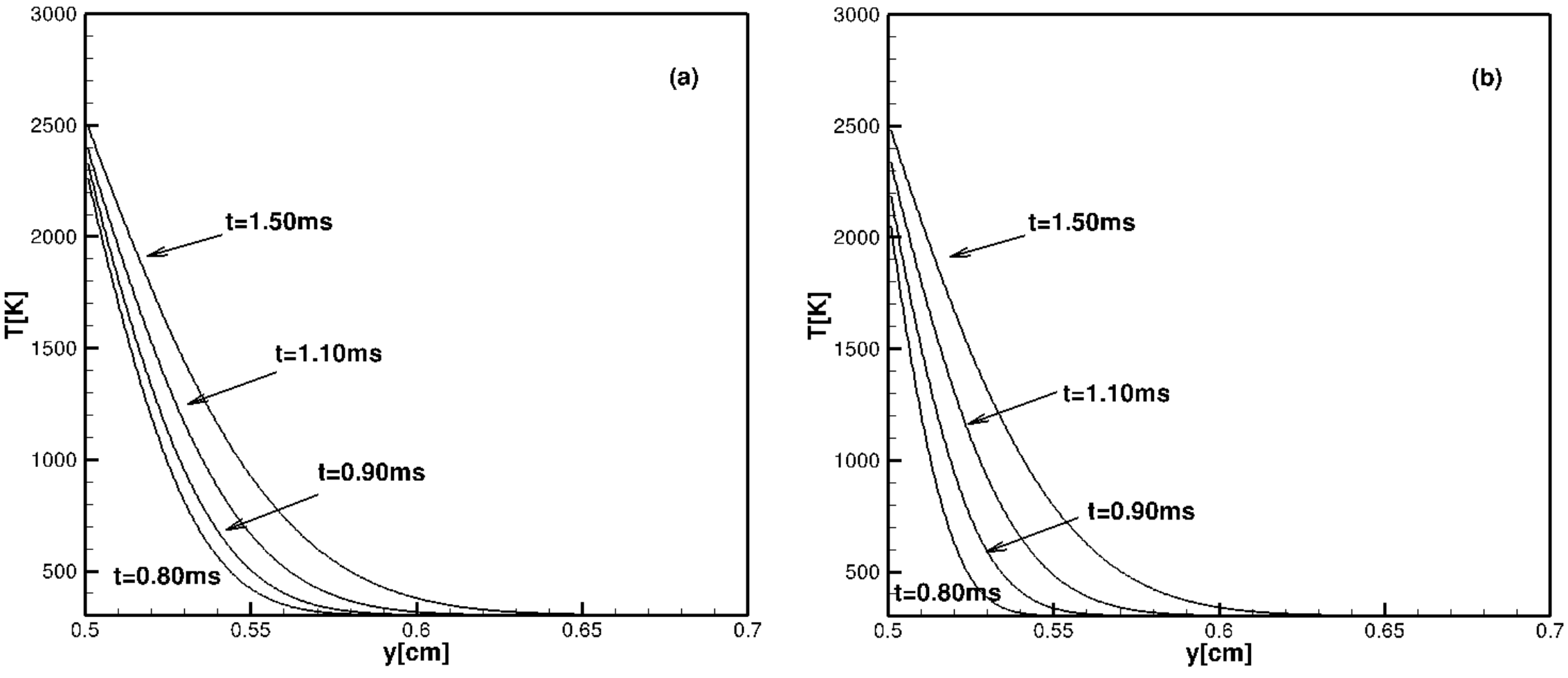

Fig. 9(a, b) Temperature profiles inside the copper walls. Parameters are the same as in Fig.8 (a,b).

It can be seen that, for both quartz and copper walls, the temperature at the inner wall surface varies significantly with time, in contrast to models employing isothermal boundary conditions. The heat penetration depth associated with the temperature profiles shown in Figs. 8(a,b) and 9(a,b) is in good agreement with the well-known one-dimensional analytical solution, $\Delta y \propto 2\sqrt{\chi t}$. The significantly higher thermal diffusivity of copper compared with quartz results in higher temperatures at the inner wall surface near the left closed end at (t = 0.8) ms and (t = 1.5) ms, as shown in Fig. 9(a), than those obtained for the quartz wall in Fig. 8(a). It should be emphasized that, although heat loss has only a minor influence on flame dynamics during the early stage of flame propagation, its effect may become much more pronounced at later stages, such as during the deflagration-to-detonation transition (DDT). Since the DDT process typically develops over a much longer time scale, on the order of 5–10 ms depending on the tube diameter, substantially greater heat losses can accumulate.

At sufficiently high Reynolds numbers, the convective heat transfer coefficient increases as $h_i \propto \sqrt{\mathrm{Re}}$, while the thermal boundary layer thickness decreases as $\delta \propto 1/\sqrt{\mathrm{Re}}$. Consequently, heat is removed more efficiently from the near-wall gas, resulting in stronger cooling of the boundary layer and a lower temperature at the inner wall surface. This behavior is illustrated by the time evolution of the temperature profiles shown in Figs. 10(a,b), computed using the convective heat transfer coefficient $h_i = 400\,\mathrm{W/m^2K}$. The results are presented for channels with quartz walls in Fig. 10(a) and copper walls in Fig. 10(b).

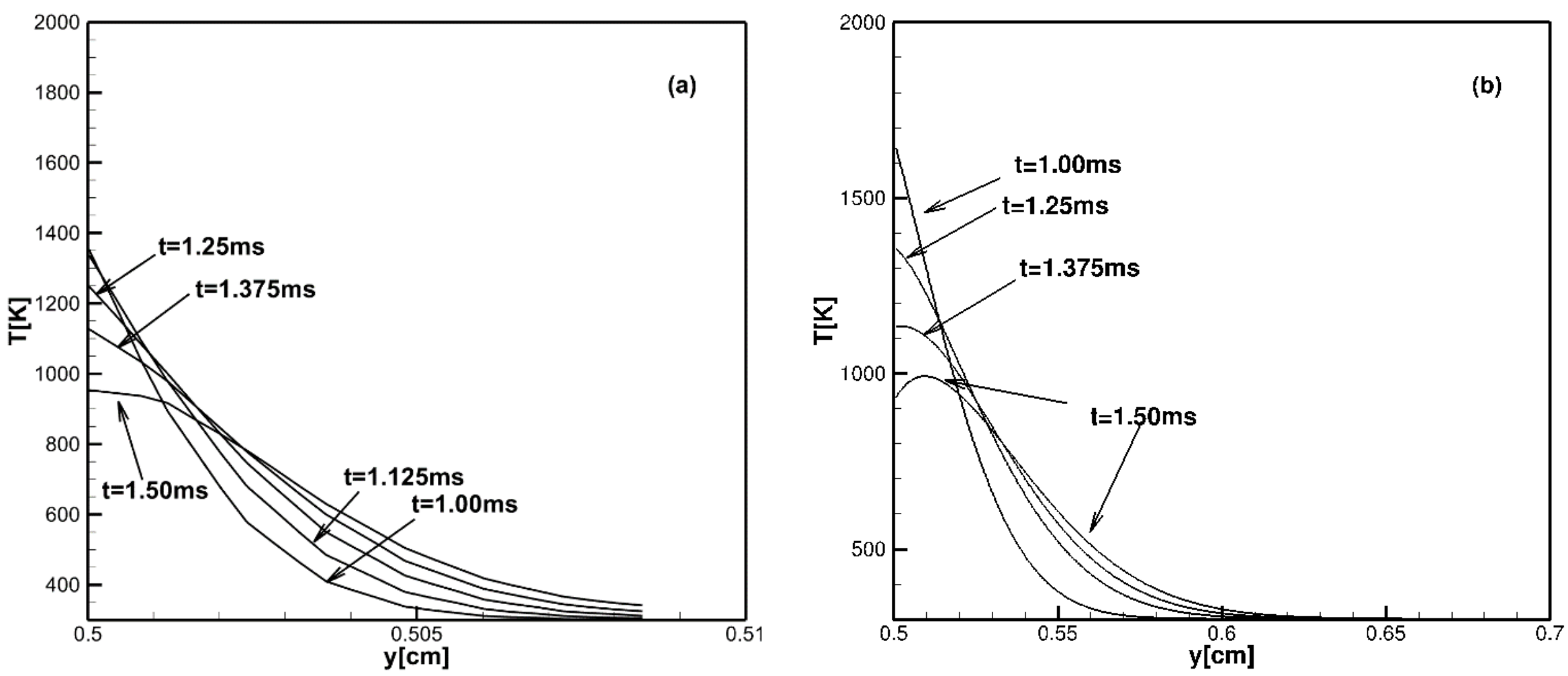

Fig. 10(a, b) Temperature profiles inside the channel wall at x=1.5cm calculated for $h_i = 400\,\mathrm{W/m^2K}$: (a) quartz wall; (b) copper walls.

The effect of wall thermal conductivity is likely to become significant only for very thin walls, on the order of $d \approx 0.1\,\mathrm{mm}$. While a high convective heat transfer coefficient can strongly influence the flame-front inversion time. This is illustrated in Fig. 11(a, b), which shows results of calculations of $H_2$/air flame propagating in a two-dimensional channel with an aspect ratio $L/D = 6$, and quartz walls of thickness $d = 2\,\text{mm}$. The figures present velocity of the flame front along the channel axis at $y = 0$ and near the wall at $y = 0.42cm$ for $h_i = 100\,\text{W/m}^2\text{K}$ shown in Fig. 11(a) and for $h_i = 400\,\text{W/m}^2\text{K}$ shown in Fig. 11(b), respectively. It is seen that at the larger heat transfer coefficient the flame front inversion and tulip flame formation occur earlier, because at a larger convective heat transfer the flame front velocity along the axis decreases faster.

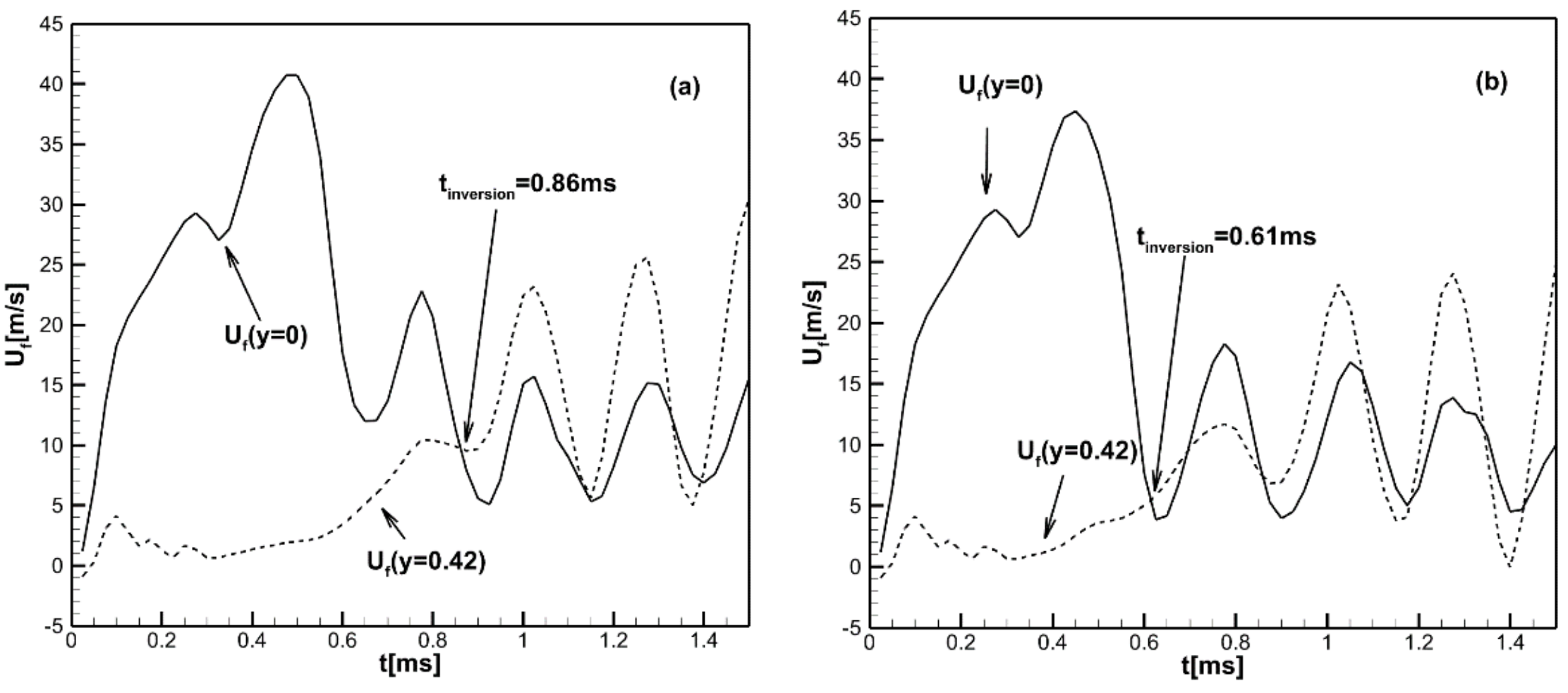

Fig. 11(a,b). Calculated flame front velocities along the channel axis and near the channel wall: (a) $h_i = 100\,\text{W/m}^2\text{K}$, (b) $h_i = 400\,\text{W/m}^2\text{K}$.

## 4. Flame in cylindrical tube

In our previous studies [25–27], we demonstrated that flame front inversion, tulip flame formation, and the subsequent evolution of the flame shape occur significantly faster in three-dimensional flames than in two-dimensional flames. To understand this difference, it is necessary to clarify the meaning of the term "flame speed," since different points on a curved (wavy) flame front propagate at different local speeds.

The propagation velocity of a flame, or more precisely the combustion wave velocity, is determined by the average fuel consumption rate, which is proportional to the total heat release rate per unit projected frontal area of the flame. Because a wrinkled flame has a larger surface area than a planar flame, it consumes fuel at a higher overall rate, resulting in a higher combustion wave velocity. This velocity, in turn, determines the flow velocity of the unburned mixture ahead of the flame. The different evolution of the combustion wave velocity in two- and three-dimensional flames arises from their fundamentally different geometries. In two dimensions, the flame front is a line, whereas in three dimensions it is a surface. Consequently, the same degree of flame wrinkling or curvature produces much larger changes in flame surface area in three dimensions than in two dimensions. This results in larger variations in the total heat release rate and, consequently, in the combustion wave velocity.

In this section, we investigate the propagation of an axisymmetric flame in a cylindrical tube with a diameter $D_{in} = 1\text{cm}$ equal to the inner width of the two-dimensional channels considered above. An axisymmetric flame in a cylindrical tube provides an intermediate model between two-dimensional and fully three-dimensional flames. It captures the key characteristics of three-dimensional flow while requiring substantially less computational resources than fully three-dimensional simulations.

Figure 12(a) shows the evolution of the flame front velocities along the tube axis, $U_f(r=0)$ and near the tube wall, $U_f(r=0.42\,\text{cm})$, for the axisymmetric flame in a cylindrical tube (x,r), together with the corresponding velocities of a two-dimensional flame propagating in a channel of the same width, $D_{in} = 1\text{cm}$. Figure 12(b) compares the corresponding flame front velocities for the axisymmetric flame with those of a fully three-dimensional flame propagating in a channel with a square cross-section of $1 \times 1\text{cm}^2$. All calculations presented in Fig. 12(a,b) were performed under adiabatic wall boundary conditions for flames propagating in channals with both ends closed and an aspect ratio $L / D_{in} = 6$. It is seen that flame front inversion takes the longest time for 2D flames (0.85ms), for cylindrical flame (0.55ms) and for 3D flame it is 0.50ms.

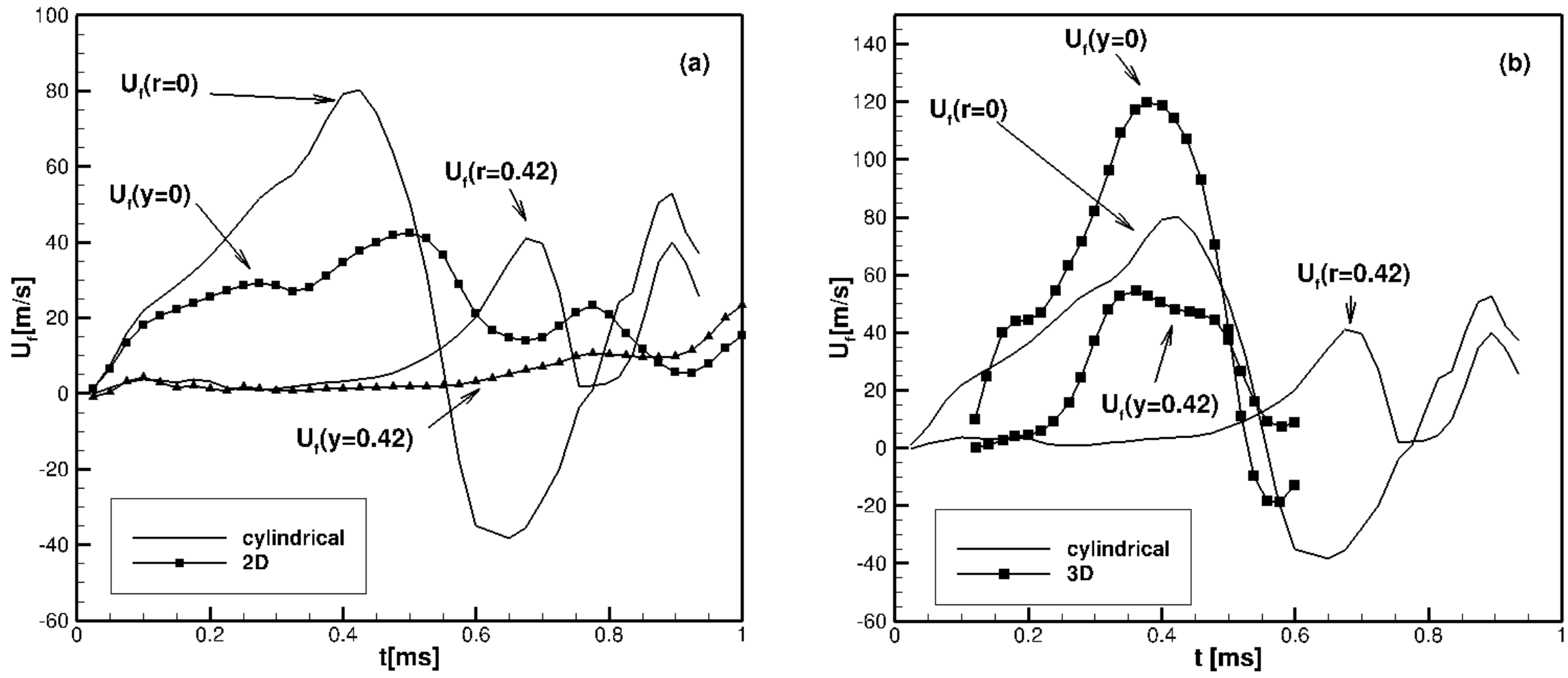

Fig. 12(a, b). (a): The flame front velocities along the axis and near the wall, during tulip flame formation in axisymmetric cylindrical flame and corresponding flame front velocities in two-dimensional channel of width $D_{in}$; (b): the calculated flame front velocities along the axis and near the wall for axisymmetric cylindrical flame and for the flame propagating in the three dimensional channel of the same width.

Fig. 13 shows the calculated the time evolution of the axisymmetric cylindrical flame front velocity along the tube axis, at $r = 0$, and at $r = 0.42\,\text{cm}$ for adiabatic (solid lines) and heat loss (lines with squares and triangles) boundary conditions in the cylindrical tube with both ends closed, an aspect ratio $\alpha = 6$, $h_i = 100\,\text{W/m}^2\text{K}$, quartz wall thickness of $d = 2\,\text{mm}$.

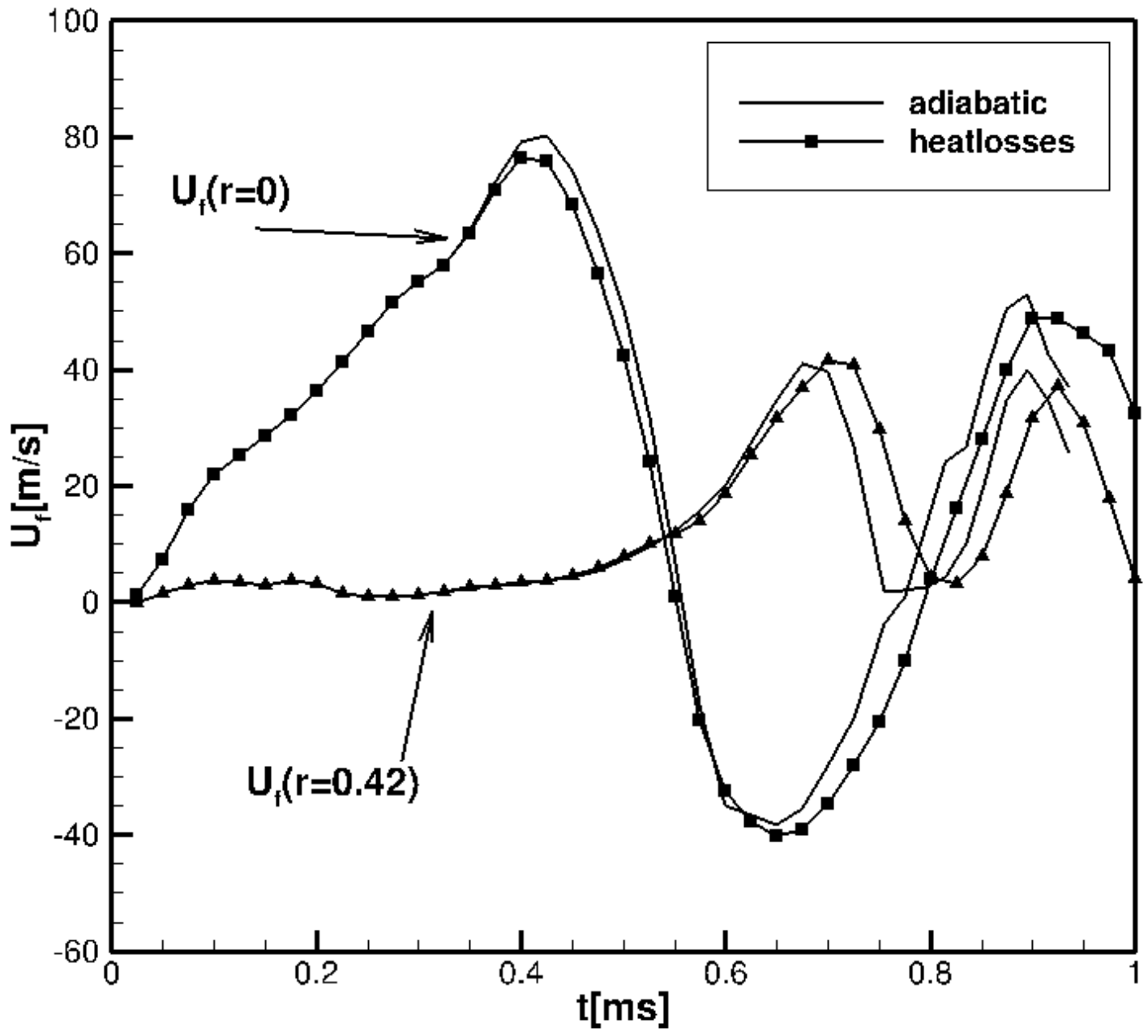

Fig. 13. Evolution of axysimmetric $H_2$/air flame front velocities for adiabatic (solid lines) and heat loss (marked lines) boundary condition; $\alpha = 6$, $h_i = 100\,\text{W/m}^2\text{K}$, quartz wall $d = 2\,\text{mm}$.

One might expect that, because the tulip flame forms more rapidly in a cylindrical tube than in a two-dimensional channel, heat loss would have a weaker influence on the flame dynamics. However, Fig. 13 shows that the effect of heat loss is, in fact, even more pronounced for the cylindrical flame than for the two-dimensional flame shown in Fig. 2. Although the characteristic times associated with the initial flame propagation and flame front inversion are shorter for the axisymmetric flame, the larger surface-to-volume ratio in the cylindrical tube results in greater heat losses, leading to a greater reduction in the average gas temperature and, consequently, in the pressure, which outweighs the effect of the shorter propagation time. For a real 3D flame the effect of heat loss would have a weaker influence on the flame dynamics.

During the early stage of flame propagation, the differences between the flame velocities computed under adiabatic and heat-loss boundary conditions are relatively small for both cylindrical and two-dimensional flames. However, for channels with both ends closed and having the same width (diameter) and aspect ratio, the pressure rises much more rapidly in the cylindrical tube because the axisymmetric flame accelerates more strongly than the two-dimensional flame.

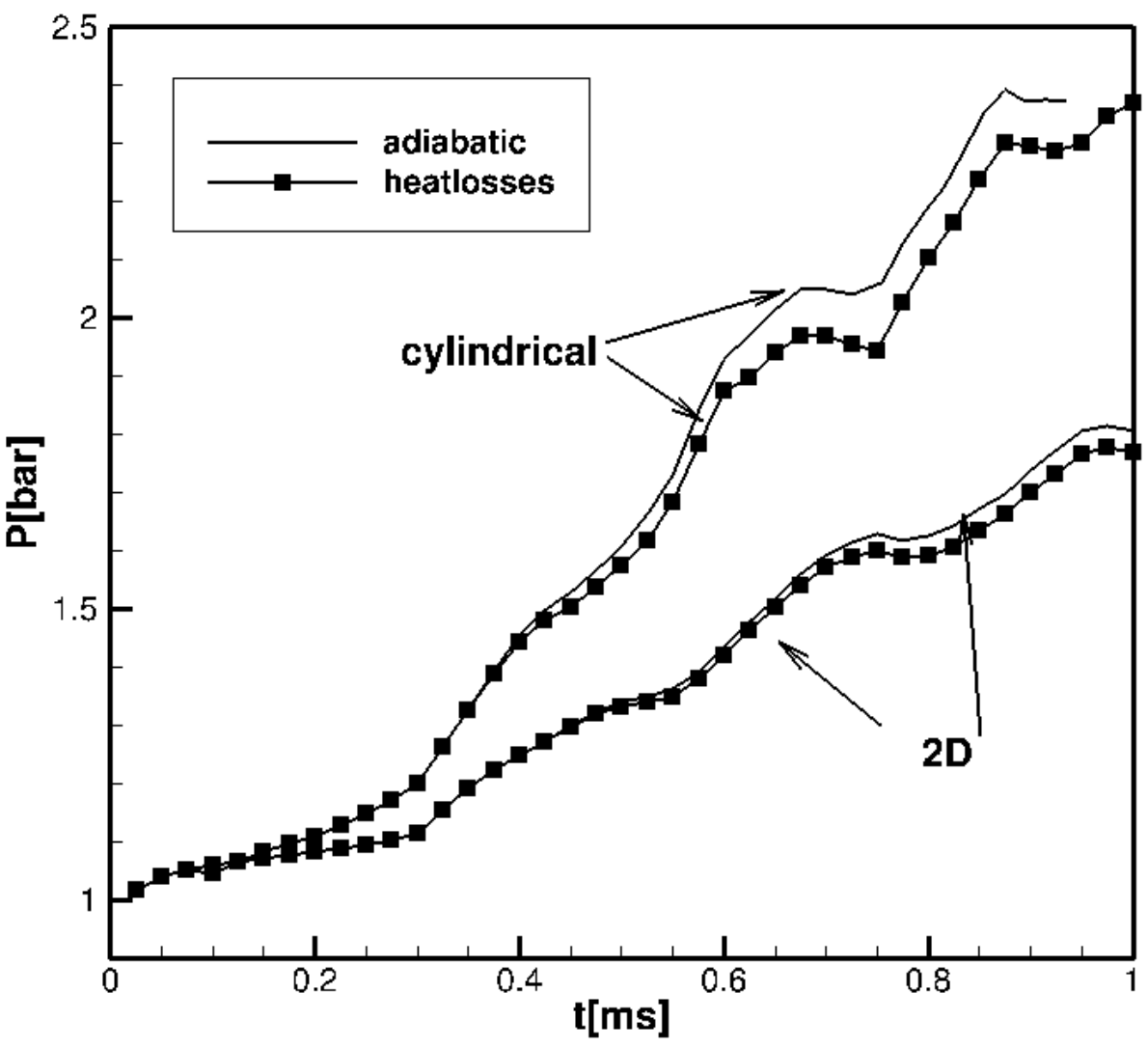

Fig.14. Evolution of pressure near the left end for adiabatic and heat losse boundary conditions for cylindrical and 2D flames. $L/D_{in} = 6$, $h_i = 100\,\mathrm{W/m^2K}$, quartz wall thickness $d = 2\,\mathrm{mm}$.

Figure 14 compares the pressure evolution near the left closed end for both the cylindrical and 2D flames under adiabatic and heat-loss boundary conditions. It can be seen that, although the pressure waves generated by the faster cylindrical flame are stronger than those produced by the two-dimensional flame, heat losses suppress the pressure rise more effectively in the cylindrical tube than in the two-dimensional channel.

## 5. Discussion and conclusions:

The effect of heat loss on early stages of flame propagation in two-dimensional channels and in cylindrical tubes were performed considering a complete model of the convective heat transfer from the hot combustion products to the inner surface of the channel walls, the thermal conductive heat propagation from the inner to the outer surface of the wall and convective and radiative heat loss from the outer surface to the surroundings. Different scenarios of heat loss were considered depending on thermal conductivity of the wall and intencity of the convective heat flux - the value of convective heat-transfer coefficient. Comparision of the flame propagation in 2D channels and in a cylindrical tube has shown that effect of heat lossss is enhanced with the increase of the ratio of the wall surface area occupied by the burnt gas to the total volume of hot combustion products. It was found that the effct of heat loss is mostly affected be the time of a flame propagation and by the ratio of the wall surface area the volume of hot combustion products. It was shown that temperature of inner walls changes considerably during the early stages of flame propagation, which means that commonly used simple models with cold isothermal walls do not adecvately describe the actual effect of heat loss.

In practice, the noticible influence of heat loss on flame acceleration is critical if the flame propagates in a low reactive mixture so that its velocity is sufficiently small. Adiabatic boundary conditions suppress the heat necessarily lost by the flame and promote the flame acceleration. Isothermal boundary conditions are valid only if the flame speed is very high such that walls do not have time to heat up and their temperature remains equal to the initial temperature. A real experimental scenario lies between these two boundary conditions, and therefore both of them either underestimate or overestimate the flame acceleration.

For a relatively thick channel walls (d=2mm), the heat transferred by thermal conduction in solid walls has no time to reach the outer wall surface during early stages of flame propagation even for wall material with high thermal conductivity like a copper. The temperature of the outer wall surface would noticibly increase during early stages only for the wall thickness less than 0.01mm for quartz walls or less than 0.1mm for copper walls, which are unrealistic practically. Therefore, radiative and convective heat losses from the outer wall to the surrounds are not important on early stages of flame propagation.

It is interesting to consider regimes for the limiting values of parameters $h_i$, $\lambda_w$, $h_{ext}$. In the case of $\lambda_w \to 0$, the flux of heat in the wall vanishes and an adiabatic flame develops whatever values of $h_i$ and $h_{ext}$. For very large values of wall conductivity $\lambda_w \to \infty$, corresponds to isothermal conditions with an almost uniform temperature in the wall, which rises in time whatever values $h_i$ and $h_{ext}$. For the very large value of $h_{ext}$, the wall becomes isothermal, whatever values $h_i$ and $\lambda_w$. For the very small value of $h_{ext} \to 0$, the outside surface of the wall is insulated and temperature of the wall increases with time whatever the values of $h_i$ and $\lambda_w$. In the limiting case of a very small $h_i$ the heat flux to the wall vanishes, which corresponds to adiabatic boundary conditions, regardless of the values of $\lambda_w$ and $h_{ext}$. The limiting case of a very large $h_i$ corresponds to isothermal walls, but practically unrealistic. According to Eqs.(20, 21) this can be achived only for a very large velocity of the burned gas, $U_b$, and a small channel width, $D_{in}$.

It is interesting to note that the tulip flame formation have been observed in 2D and 3D numerical studies of hydrogen-air flames dynamics in heat-recirculating microtubes [40, 42 – 45], where the flame is stabilized by an inlet mass flow rate matching the flame burning velocity. The concept of micro-combustors relies on heat transfer through solid walls, which provide an additional pathway for heat to reach the unburned gas ahead of the flame front, supplementing the conventional gas-phase thermal conduction [71–73]. For freely propagating flame this is impossible because the rate of the heat transfer by gas-phase thermal conduction is much faster than heat transfer through solid materials. The heat recirculation is possible only if the flame is stabilized by the flammability limit in narrow tubes or by a counterflow of fresh gas. It should be emphasized that, unlike a freely propagating flame in a tube, where the mechanism of tulip flame formation is a purely hydrodynamic phenomenon [25, 29], the inversion of the flame front observed in [40, 42 – 45] is a purely thermal phenomenon. In the heat-recirculating tubes heat from the hot burned gas is transferred to tube walls and by the wall thermal conduction transfers in the axial direction to preheat the unburned gas near the wall ahead of the flame front. Therefore, parts of the flame front near the wall will propagate faster than the parts of the flame front near the tube axis, where the flame front velocity will be minimum.

## 6. Appendix A

Figures A1(a, b) show the time evolution of the flame front velocities along the channel axis, $U_f(y=0)$, and the flame front velocity close to sidewall, $U_f(y=0.42cm)$ obtained in simulations with a one-step chemical model for hydrogen/air, Fig. A1(a), and in simulations with a detailed chemistry, Fig. A1(b).

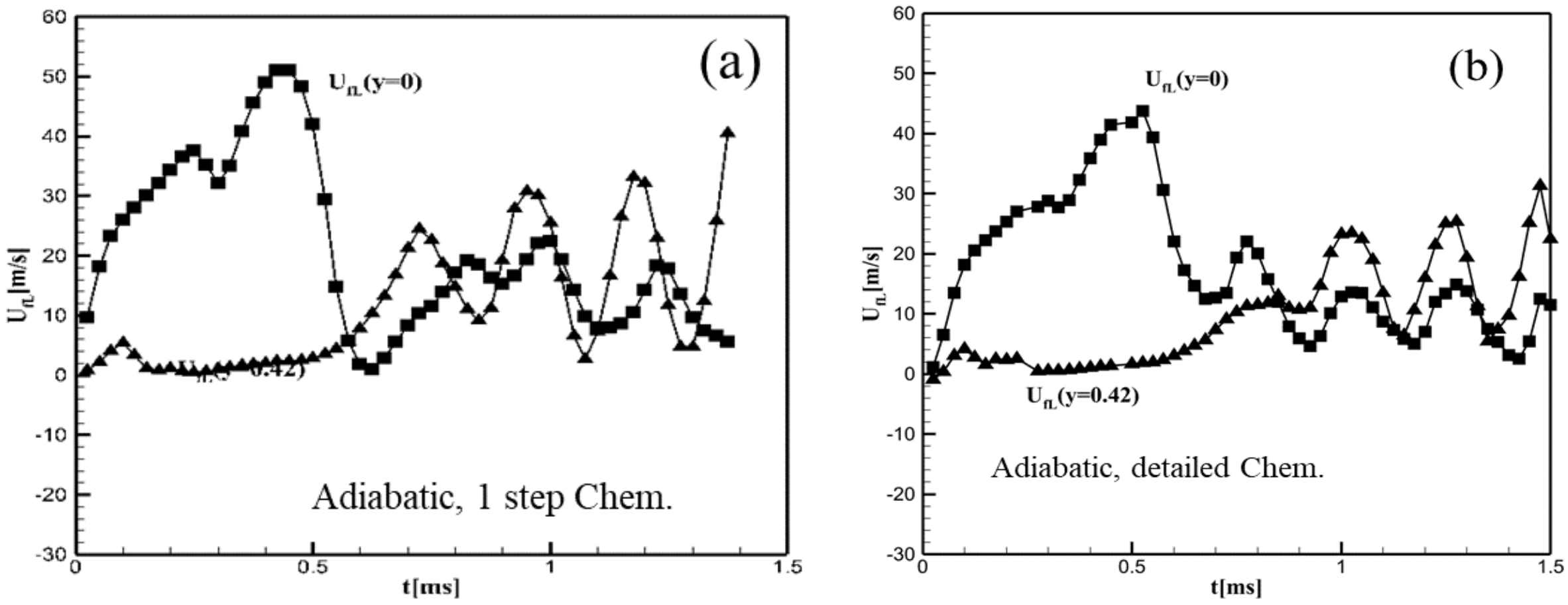

Fig. A1(a, b). (a): Time evolution of the hydrogen-air flame front velocities along the tube axis $U_f(y=0)$ and near the wall $U_f(y=0.42cm)$ obtained in simulations with adiabatic boundary conditions and a single-step chemical model; (b): the same but for a detailed chemical model.

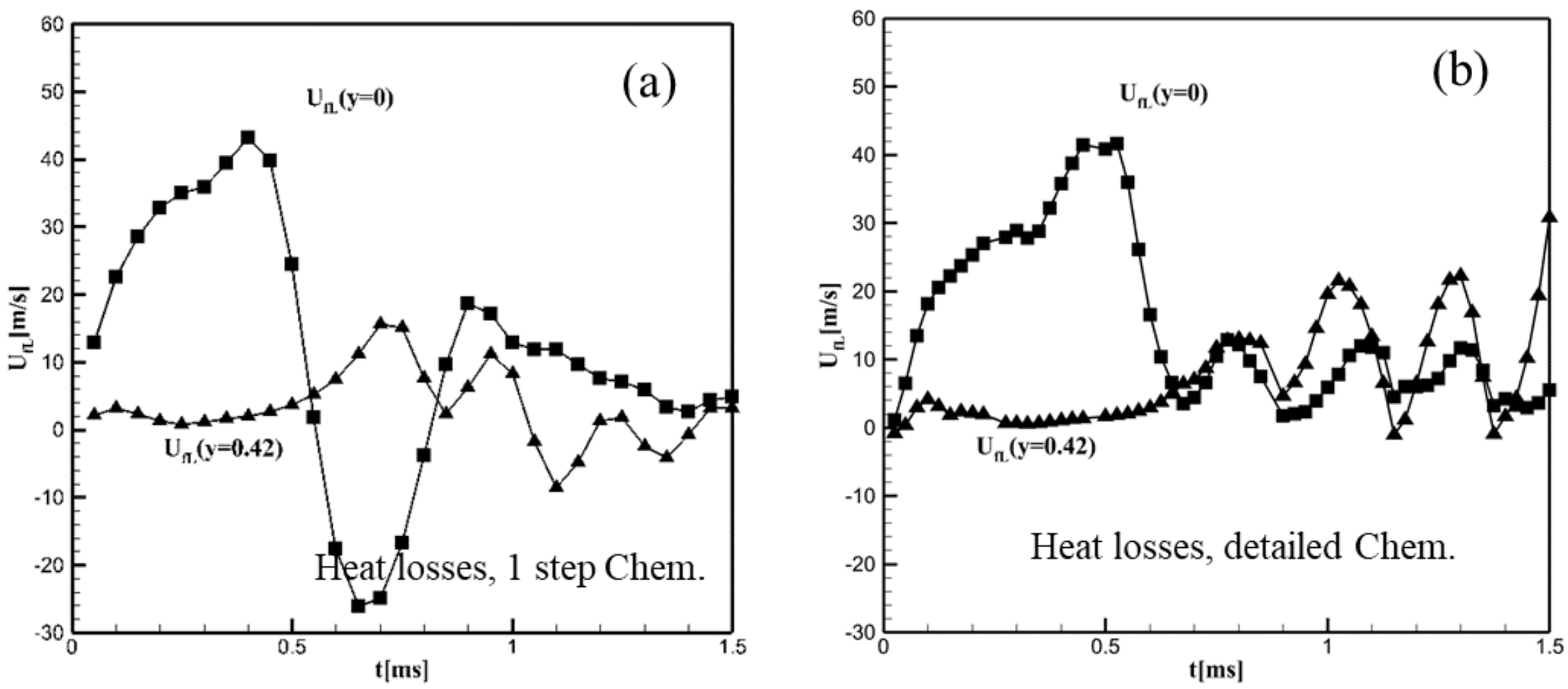

Fig.A2(a, b). (a): Time evolution of the hydrogen-air flame front velocities along the tube axis and near the wall, $y=4.2mm$, obtained in simulations with with heat losses to the walls using a one-step chemical model; (b) the same but for a detailed chemical model.

The flame dynamics obtained in simulations using either adiabatic or heat loss boundary conditions are considerably different for simulations with a one-step chemical model compared to simulations with a detailed chemical kinetics, as it is seen in Fig. A1(a, b) and Fig. A2(a, b).

**Novelty and significance statement**

This study presents the first rigorous and systematic investigation of heat-loss effects on flame shape and acceleration during the early stages of flame propagation. The analysis accounts for convective heat transfer from hot combustion products to the wall, thermal conductivity heat transfer to the outer wall surface, and convective and radiative heat losses from the outer wall surface to the surroundings. The study examines hydrogen–air in two-dimensional channels with different aspect ratios, and axisymmetric hydrogen–air flames in cylindrical tubes. By identifying how heat loss influences flame structure and acceleration, this work advances the fundamental understanding of flame acceleration and provides insights essential for predicting and controlling combustion in practical systems.

**CRediT authorship contribution statement**

Michael Liberman: problem set-up & conceptualization, writing—original draft preparation, review & editing, supervision, methodology. Chengeng Qian: software, visualization, investigation, formal analysis, data curation, writing – review & editing.

**Declaration of competing interest**

The authors declare that they have no known competing financial interests or personal relationships that could have appeared to influence the work reported in this paper.

**Acknowledgements**

M.L. greatly acknowledges the financial support of this research provided by the Olle Engkvists Stiftelse (Foundation) under grant No.232-0212, project No 31005258.